\documentclass[12pt]{iopart}
\usepackage{graphicx}
\usepackage{epsfig}
\usepackage{ulem}

\begin{document}

\title[Dynamics of one-dimensional tight-binding models with
arbitrary time-dependent]{Dynamics of one-dimensional tight-binding
models with arbitrary time-dependent external homogeneous fields}
\author{W H Hu, L Jin and Z Song}
\address{School of Physics, Nankai University, Tianjin 300071,
China}
\ead{songtc@nankai.edu.cn}

\begin{abstract}
The exact propagators of two one-dimensional systems with
time-dependent external fields are presented by following the
path-integral method. It is shown that the Bloch acceleration theorem
can be generalized to the impulse-momentum theorem in a quantum
version. We demonstrate that an evolved Gaussian wave packet always
retains its shape in an arbitrary time-dependent homogeneous driven
field. Moreover, the stopping and accelerating of a wave packet can
be achieved by a pulsed field in a diabatic way.
\end{abstract}


\maketitle

\section{Introduction}
\label{sec.intro}

The possibility of dynamically controlling the quantum state by a
time-dependent external field represents a promising and powerful
approach for quantum devices. Dynamical decoupling (DD) is a paradigm
of bang-bang control techniques that applies a specially designed
sequence of control pulses to the qubits to negate the coupling of
the central system to their environment \cite{DD}. Moreover, the
quantum Zeno effect has been proposed as a strategy to protect the
coherence \cite{DD1,DD2}. Recently, it has been proposed that a
periodically driven potential can suppress the tunnelling between
adjacent sites in a lattice \cite{OPL1,OPL2}. It is remarkable that,
although the scheme for the purpose has been studied extensively in
various systems, a proposal has not yet been made on the basis of an
\textit{aperiodic} external field.

In this study, in general, we will generally study the influence of
arbitrary time-dependent electric and magnetic fields on the dynamics
of a quantum state in one-dimensional tight-binding systems. Our
analysis is focused on the propagator, which governs the time
evolution of an arbitrary quantum state. We consider two related
models, an infinite tight-binding chain subjected to an arbitrary
time-dependent linear potential and a finite ring threaded by an
arbitrary time-dependent flux that allow analytical treatments of the
dynamics of matter waves and provide a valuable insight into it. We
will show that a temporal modulation of the external field, which is
subjected to the infinite chain or the finite ring, can be exploited
to coherently and reversibly control the wave vector and the phase of
the matter wave. It offers the advantage that the dynamical external
field is not necessarily required to maintain an adiabatic change. As
an application, this feature can be used to stop and accelerate a
wave packet or transfer it to any location on demand.

In Sec.~\ref{sec.models}, we derive analytical expressions for the
propagator of two such time-dependent systems, thus completing the
description of the controllability of the system for a quantum state.
Sec.~\ref{sec.dyn} is devoted to the application of the propagator to
several specialized cases. These include the detailed treatments and
computations of the time evolutions of a typical initial state under
time-periodic external fields. In Sec.~\ref{sec.QST}, a scheme is
presented for quantum state control and manipulation. Final
conclusions and discussions are drawn in Sec.~\ref{sec.Sum}.

\section{Models and propagators}
\label{sec.models}

In this section, we present charged particle models under
considerations, the simple tight-binding model in external electric
and magnetic fields. Here, the transition between Bloch bands and
particle-particle interaction is ignored for simplicity. This
approach is based on our previous study \cite{YS1,Hu_AC}, where we
proposed a scheme for quantum state manipulation. It employed a loop
enclosing a magnetic flux to control the speed of a wave packet. We
have shown how to move a Gaussian wave packet (GWP) at a certain
speed on demand by a static magnetic flux and freeze it by a
periodically alternating flux. These are the basic operations for
quantum-state engineering and quantum information processing (QIP).
On the other hand, another scheme for this purpose that exploits a
linear potential with sinusoidal time modulation has attracted
attention \cite{PRL_PD,PRA_SBO}. In this study, we aim at employing
the same systems with \textit{aperiodically} time-dependent fields
for the coherent control of matter waves. In such schemes, one can
accomplish the task of accelerating and stopping a wave packet by a
pulsed field. The central object in studying the dynamics of a
quantum system is the propagator, which is the transition amplitude
between the states at the initial and final instants in time and
which governs the time evolution of any matter wave. In the
following, we will systematically address the dynamics of these two
systems by deriving analytical propagators.

\begin{figure}[tbp]
\begin{center}
\includegraphics[bb=0 0 540 380,width=0.60\textwidth,clip]%
{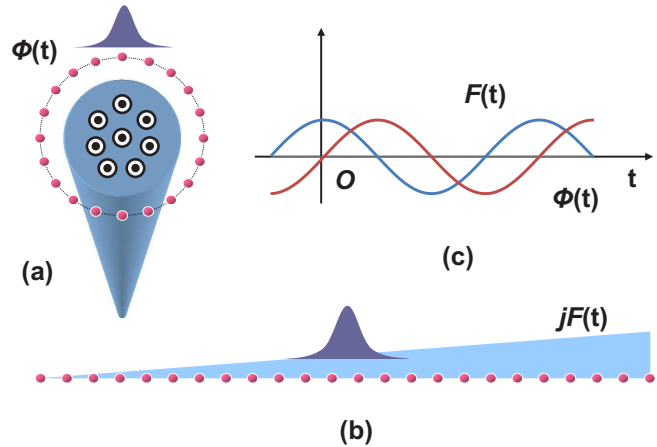}
\caption{(Color online) The schematic illustrations for a
tight-binding ring threaded by a time-dependent magnetic flux
$\Phi \left( t \right)$ in (a) from (\ref{HB}), and a chain
subjected to a time-dependent external field $F \left( t \right)$ in
(b) from (\ref{HE}), where $j$ denotes the lattice site. The flux
and the external field is illustrated in (c), which are, but not
necessary, time-periodic.}
\label{fig.illus}
\end{center}
\end{figure}

\subsection{Finite ring}
\label{sec.mod.Ring}

Consider a ring lattice with $N$ sites threaded by a magnetic field,
as illustrated schematically in Figure~\ref{fig.illus}(a). The
Hamiltonian of the corresponding tight-binding model
\begin{equation}
H_{\mathrm{B}} = -J \sum_{j=1}^{N} \left(  \rme ^{ \rmi 2\pi \Phi \left(
t \right) /N}a_{j}^{\dagger} a_{j+1} + \mathrm{H.c.}\right)
\label{HB}
\end{equation}
depends on the magnetic flux through the ring in units of the
magnetic flux quantum $\Phi_{0} = h/e$. Here $a_{j}^{\dagger}$ is the
creation operator of a particle at the $j$th site with periodic
boundary conditions. The flux does not exert force on the Bloch
electrons but can change the local phase of its wave function due to
the Aharonov-Bohm (AB) effect. Note that the particle is not
restricted to be either bosons or fermions.

By the transformation
\begin{equation}
a_{j} = \frac{1}{\sqrt{N}} \sum_{k}  \rme ^{ \rmi kj} a_{k},
\label{a_j}
\end{equation}
where $k=2\pi n/N$ and $n \in \left[ 1,N\right] $, the Hamiltonian
can readily be written as
\begin{equation}
H_{\mathrm{B}} = -2J \sum_{k} \cos \left( k + \phi \left( t \right)
\right) a_{k}^{\dagger} a_{k},  \label{HB_k}
\end{equation}
with $\phi \left( t \right) = 2\pi \Phi \left( t\right) /N$, and the
corresponding eigenstate takes the form of
\begin{equation}
\left\vert k \right\rangle = \frac{1}{\sqrt{N}} \sum_{j}  \rme ^{
\rmi kj} \left\vert j \right\rangle .  \label{a_k}
\end{equation}
Note that the time-dependent Hamiltonian possesses fixed
instantaneous eigenstates, while the flux solely affects the
instantaneous eigenvalues. This will be crucial for employing such
a setup to investigate the control of a quantum state since the exact
solutions of a time-dependent Hamiltonian are rare.

The evolution of an arbitrary state under the Hamiltonian $H$ is
dictated by the unitary operator
\begin{equation}
U \left( t^{\prime },t\right) =\mathcal{T} \exp \left( -i
\int_{t}^{t^{\prime }} H \left( t^{\prime \prime} \right) \rmd
t^{\prime \prime } \right) ,
\end{equation}
where $\mathcal{T}$ denotes the time-ordering operator, which yields
the propagator represented in the momentum and the spatial
eigenstates
\begin{eqnarray}
U_{k^{\prime }k}\left( t^{\prime },t\right)
=& \left\langle k^{\prime} \right\vert U \left( t^{\prime },t\right)
\left\vert k \right\rangle &=  \exp \left[ \rmi 2J f_{k} \left( t^{\prime } , t
\right) \right] \delta_{k^{\prime }k},  \\
U_{j^{\prime }j}\left( t^{\prime },t\right)
=& \left\langle j^{\prime} \right\vert U \left( t^{\prime},t\right)
\left\vert j \right\rangle &= \frac{1}{N} \sum_{k} \exp \left[ \rmi
2J f_{k} \left( t^{\prime}, t\right) \right] \rme ^{ \rmi k\left(
j^{\prime } - j \right)}  ,
\end{eqnarray}
where
\begin{eqnarray}
f_{k} \left( t^{\prime },t\right)
&=& \int_{t}^{t^{\prime}} \cos \left[ k+ \phi \left( t^{\prime \prime
} \right) \right] \rmd t^{\prime \prime } = \cos \left(k \right) u +
\sin \left(k \right) v \nonumber \\
&=& \sqrt{u^{2}+v^{2}} \cos \left[ k + \varphi \left( t^{\prime }, t
\right) \right] .
\end{eqnarray}
Here the time-dependent functions are defined as
\begin{eqnarray}
\varphi \left( t^{\prime },t\right) &=& \arg \left( u-iv\right) ,\\
\qquad u &=& \int_{t}^{t^{\prime}} \cos \left[ \phi \left( t^{\prime
\prime} \right) \right] \rmd t^{\prime \prime},  \\
\qquad v &=& \int_{t}^{t^{\prime}} \sin \left[ \phi \left( t^{\prime
\prime} \right) \right] \rmd t^{\prime \prime}.
\end{eqnarray}
Thus, the propagator can be obtained in the following explicit form
\begin{equation}
U_{k^{\prime }k} \left( t^{\prime}, t\right)
= \exp \left( \rmi 2J_{\mathrm{eff}} \cos \left[ k + \varphi \left(
t^{\prime}, t \right) \right] \left( t^{\prime} - t \right) \right)
\delta \left( k^{\prime} - k\right) \label{eq.U_ring}
\end{equation}
with
\begin{equation*}
J_{\mathrm{eff}} = J\sqrt{u^{2}+v^{2}} / \left( t^{\prime} -t\right).
\end{equation*}
Its physical interpretation is very clear: the time evolution of the
state at instant $t^{\prime}$ is equivalent to that governed by a
uniform ring with hopping strength $J_{\mathrm{eff}}$ threaded by the
flux $N \varphi \left( t^{\prime},t\right) $. We note that such an
equivalence is always true for an arbitrary time-dependent function
of the flux $\phi \left( t \right)$. In the following, we will
discuss its application together with the propagator of the infinite
chain system.

\subsection{Infinite chain}
\label{sec.mod.Chain}

Now we turn to an infinite chain system driven by a time-dependent
homogeneous field, as illustrated schematically in
Figure~\ref{fig.illus}(b). The Hamiltonian of the corresponding
tight-binding model is
\begin{equation}
H_{\mathrm{E}} = -J \sum_{j=-\infty}^{\infty} \left( a_{j}^{\dagger}
a_{j+1} + \mathrm{H.c.} \right) + F\left( t \right) \sum
_{j = -\infty} ^{\infty} j n_{j} . \label{HE}
\end{equation}
In the case of a constant external field $F_{0}$, the dynamics of a
localized wave packet exhibit coherent oscillations with period
$\tau _{\mathrm{BO}}=2\pi /F_{0}$, i.e. Bloch oscillations, instead
of the expected accelerated motion towards infinity \cite{Fukuyama}.
This non-intuitive feature arises from the discreteness of the
system.

Over the past several decades, such a system has been investigated
extensively for potential applications. Most of them focus on the
static or time-periodic fields. In this study, we investigate the
case of arbitrary time-dependent homogeneous fields.

Unlike the system of (\ref{HB}) discussed above, the instantaneous
eigenfunctions are no longer independent of time. However, they can
be obtained explicitly on the basis of the Stark ladder theory
\cite{NJP_SL}. For the time-dependent Hamiltonian, according to the
framework of Feynman's polygonal path approach, the propagator can be
presented as a functional integration
\begin{equation}
U_{k^{\prime } k } \left( t^{\prime }, t \right) = \int_{k}
^{k^{\prime}} \exp \left(- \rmi \int_{t}^{t^{\prime}} H \left( k^{\prime
\prime}, t^{\prime \prime} \right) \rmd t^{\prime \prime}
\right) \mathcal{D} \left( k^{\prime \prime} \right) ,
\end{equation}
where the integration denotes the time evolution through different
paths from $k$ to $k^{\prime}$ in the momentum space \cite{JST_Bal}.

For an infinitesimal time interval $\rmd t$, the propagator can be
obtained from that of the time-independent Hamiltonian, which has the
form
\begin{eqnarray}
U_{k^{\prime }k}\left( t+\rmd t,t\right)
= \exp \left(- \rmi 2J\frac{\sin \left( k^{\prime} \right) - \sin
\left( k \right)}{F\left( t \right)}    \right) \delta \left(
k^{\prime} - k + F \left( t \right) \rmd t \right) . \label{eq.U_div}
\end{eqnarray}
Considering a finite time interval $t^{\prime } - t$, where $t$ and
$t^{\prime}$ denote the initial and final, respectively, the
propagator can be expressed in the form
\begin{eqnarray}
U_{k^{\prime} k} \left( t^{\prime}, t \right)
= \exp \left( \rmi 2J_{\mathrm{eff}} \cos \left[ k + \varphi \left(
t^{\prime }, t \right) \right] \left( t^{\prime} - t\right) \right)
\delta \left( k^{\prime} - k + I \left( t^{\prime}, t
\right) \right) . \label{eq.U_chain}
\end{eqnarray}
Here
\begin{equation}
J_{\mathrm{eff}} = J\sqrt{u^{2}+v^{2}} / \left( t^{\prime} - t \right)
\end{equation}
acts as the time-dependent effective hopping amplitude of a uniform
chain, and
\begin{equation}
I\left( t^{\prime }, t\right) = \int_{t}^{t^{\prime}} F \left(
t^{\prime \prime} \right) \rmd t^{\prime \prime}
\end{equation}
denotes the impulse of force during the time interval $t^{\prime} -
t$, and
\begin{equation}
\varphi \left( t^{\prime },t\right) = \arg \left( u - \rmi v \right)
\end{equation}
is the phase shift, where
\begin{eqnarray}
u &=& \int_{t}^{t^{\prime }}\cos \left[ I\left( t^{\prime \prime },
t\right) \right] \rmd t^{\prime \prime}, \\
v &=& \int_{t}^{t^{\prime }}\sin \left[ I\left( t^{\prime \prime },
t\right) \right] \rmd t^{\prime \prime}.
\end{eqnarray}

One can see that the propagators of the two systems $H_{\mathrm{B}}$
and $H_{\mathrm{E}}$ have similar form, while there are still several
differences between them: First, the allowed momentum values are
different. The former is discrete for the finite site ring, whereas
the latter must be continuous. Second, in a ring system, the momentum
is always conservative according to the factor $\delta _{k^{\prime }
k}$ in (\ref{eq.U_ring}), whereas it is not in the chain system,
suggesting that one can extend the Bloch's acceleration theorem to
the momentum-impulse theorem as
\begin{equation}
k^{\prime}=k - I \left( t^{\prime}, t \right) \label{eq.IMT}
\end{equation}
for an arbitrary $F\left( t \right)$, which is the central result of
this study. Despite the difference between the two systems, we
discuss in the next section that they are equivalent for dealing with
the dynamics of a wide wave packet if we take $F \left( t \right) = -
\partial \phi \left( t\right) /\partial t$.

\section{Wave packet dynamics}
\label{sec.dyn}

In this section, we will apply the propagator to a specific case: the
time evolution of a GWP in a system driven by an arbitrary
time-dependent external field. We will briefly rederive the basic
results for the system of cosinoidal fields based on the exact
evolved wave function, which offers various advantages over the
traditional approach. In the following, we restrict the discussion to
the GWP on the lattice, with
\begin{eqnarray}
\left\vert \psi \left( 0\right) \right\rangle
= \left\vert \Psi \left( k_{0},N_{\mathrm{A}}\right) \right\rangle
= \Lambda \sum_{k} \exp \left[ - \frac{ \left( k - k_{0} \right)
^{2}}{\alpha ^{2}} - \rmi N_{\mathrm{A}} k \right] \left\vert k
\right\rangle , \label{eq.GWP}
\end{eqnarray}
where $\Lambda$ is the normalization factor, where $k_{0}$ and
$N_{\mathrm{A}}$ denote the central momentum and position of the
initial wave packet, respectively.

Applying the propagator in (\ref{eq.U_chain}), we obtain the
evolved wave function at time $t$ as
\begin{eqnarray}
\left\vert \psi \left( t\right) \right\rangle
&=& \sum_{k} U_{k k^{\prime}} \left( t,0 \right) \left\langle
k^{\prime }\right. \left\vert \psi \left( 0\right) \right\rangle
\left\vert k \right\rangle \nonumber \\
&=& \Lambda  \rme ^{- \rmi N_{\mathrm{A}} I\left( t\right) }\sum_{k}
\exp \left( - \frac{\left( k + I \left( t \right) - k_{0} \right)^{2}
}{\alpha^{2}} \right) \nonumber \\
&& \times \exp\left( \rmi 2J_{\mathrm{eff}} \left( t \right)
\cos\left[ k + I \left( t \right) + \phi \left( t \right) \right] t -
\rmi N_{\mathrm{A}} k \right) \left\vert k \right\rangle .
\end{eqnarray}
In the case of a wide wave packet, i.e. $\alpha \ll 1$, it can be
reduced as
\begin{equation}
\left\vert \psi \left( t\right) \right\rangle
\approx \exp\left( \rmi \gamma \left( t \right) \right) \left\vert
\Psi \left( k_{0} - I \left( t \right) , N_{\mathrm{A}} + D \left( t
\right) \right) \right\rangle , \label{Psi_C}
\end{equation}
where
\begin{equation}
\gamma \left( t \right) = 2J_{\mathrm{eff}} \left( t \right) \cos
\left( k_{0} + \varphi \left( t \right) \right) t - N_{\mathrm{A}} I
\left( t \right) \label{eq.GWP.phase}
\end{equation}
is an overall phase factor, and thus can be neglected when only a
single wave packet is considered. Here
\begin{equation}
D\left( t\right) = 2J_{\mathrm{eff}} \left( t \right) \sin \left[
k_{0} + \varphi \left( t\right) \right] t \label{Dt_C}
\end{equation}
denotes the displacement of the centre of the wave packet. It is
interesting to find that the evolved state $\left\vert \psi \left(
t \right) \right\rangle $ is still a GWP with the same shape as the
initial one. The shape is retained because of the field homogeneity
and the wide GWP approximation. Nevertheless, even for a narrow wave
packet, the periodicity of the propagator ensures the revival of
shape. In any case, the displacement $D \left( t \right) $ is crucial
for characterizing the dynamics of the wave packet. However, it may
have a complex dependence on the parameters of the driving force.
Fortunately, when we deal with the group velocity $v_{\mathrm{g}}
\left( t \right) = \partial D \left( t \right) / \partial t$, the
physical picture will become clear. Actually, the straightforward
algebra shows that
\begin{equation}
v_{\mathrm{g}} \left( t \right) = 2J\sin \left[ k_{0} - I\left(
t\right) \right] , \label{eq.vg_C}
\end{equation}
which accords with the fact that the central momentum of the wave
packet $k_{0} - I\left( t \right)$ originates from the
momentum-impulse theorem in the former section. We notice that the
instantaneous group velocity is solely determined by the impulse
$I\left( t \right)$, but not the details of the field function
$F\left( t \right)$. In the following, we apply this conclusion to
some special cases to demonstrate its validity and its application
in QIP.

On the other hand, we can investigate the relationship of the two
models in (\ref{HB}) and (\ref{HE}) through the dynamics of the wave
packet. By performing the same procedure as above, we can obtain the
evolved GWP under the system of (\ref{HB}) as
\begin{equation}
\left\vert \psi \left( t\right) \right\rangle \approx  \rme ^{ \rmi \gamma
\left( t \right) } \left\vert \Psi \left( k_{0} , N_{\mathrm{A}} + D
\left( t \right) \right) \right\rangle \label{Psi_R}
\end{equation}
if the the width of the wave packet is far smaller than the length of
the ring. Here
\begin{equation}
\gamma \left( t\right) = 2J_{\mathrm{eff}} \left( t\right) \left[
\cos \left( k_{0} + \varphi \left( t \right) \right) + \sin \left(
k_{0} + \varphi \left( t \right) \right) k_{0} \right] t,
\end{equation}
and
\begin{equation}
D\left( t\right) =2J_{\mathrm{eff}}\left( t\right) \sin \left[
k_{0}+\varphi \left( t\right) \right] t. \label{Dt_R}
\end{equation}
Note that the evolved state of (\ref{Psi_C}) differs from that of
(\ref{Psi_C}) in that there is no momentum shift of the wave packet
in the flux-pierced ring. Nevertheless, its group velocity is
expressed as
\begin{equation}
v_{\mathrm{g}}\left( t \right) =2J\sin \left[ k_{0}+\phi \left( t
\right) \right] , \label{eq.vg_R}
\end{equation}
which has the same form as that in (\ref{eq.vg_C}). A comparison of
(\ref{Psi_C}) and (\ref{Psi_R}) clearly shows that by taking the
substitution of
\begin{equation}
\phi \left( t\right) =-I\left( t\right) ,
\label{eq.phiI}
\end{equation}
i.e. $F\left( t\right) = -\partial \phi \left( t\right) /\partial t$,
the probability distributions of the two evolved wave packets are
identical. Therefore, when the probability current is measured, the
two systems are equivalent within a local range. It accords with
Faraday's law of induction in classical physics, wherein a changing
magnetic flux induces an electric field. In the following, we focus
the discussions on the chain system, because the results can be
simply extended to those of the ring system.

It is worth mentioning that the group velocity $v_{\mathrm{g}}$ of
(\ref{eq.vg_R}) can be understood in the following simple manner: as
is well known, the group velocity in real space is determined by the
dispersion relation with respect to $k$. For the system
$H_{\mathrm{B}}$ of (\ref{HB}), the derivative of the instantaneous
dispersion relation in (\ref{HB_k}), $\varepsilon _{k} = -2J\cos
\left[ k +\phi \left( t\right) \right] $ can directly obtain the
group velocity expressed in (\ref{eq.vg_R}). For the system
$H_{\mathrm{E}}$\ of (\ref{HE}), we can also obtain the group
velocity from the derivative of the basal dispersion relation, which
originates from the spectrum of a uniform tight-binding chain of
$\varepsilon_{k} = -2J\cos \left( k \right)$. Meanwhile, because of
electric field effect, the momentum shifts.The central momentum of
the evolved GWP is presented by $k_{0} \left( t \right) = k_{0} - I
\left( t \right)$, which yields the group velocity of
(\ref{eq.vg_C}).

\subsection{Bloch oscillations}
\label{sec.dyn.BO}

Now, we consider the simplest case of an external static field
$F\left( t\right) =F_{0}>0$ subjected to an infinite chain system (or
equivalently, with a pierced flux $\phi \left( t\right) = -F_{0}t$ in
a finite ring system). It is well known that Bloch oscillations occur
in this system. This phenomenon can be revisited in the framework of
the propagator, which is introduced as above. Moreover, the results
can be extended to the case of a finite ring. It is also a starting
place for the discussion of more complicated cases in the next
sections.

In this case, we have
\begin{equation}
I\left( t \right) = F_{0} t.
\end{equation}
Because of the discreteness of the lattice, the momentum space is
restricted in the first Brillouin zone. Also, the Bragg reflections
give the Bloch oscillations directly. From the former analysis, we
can get
\begin{eqnarray}
D\left( t\right) &=& \frac{2J}{F_{0}}\left[ \cos \left( k_{0} - F_{0}
t \right) -\cos \left( k_{0}\right) \right] , \label{D_BO} \\
v_{\mathrm{g}}\left( t\right) &=&2J\sin \left( k_{0}-F_{0}t\right) ,
\end{eqnarray}
with the period
\begin{equation}
\tau _{\mathrm{BO}} = \frac{2\pi}{F_{0}} \label{eq.tau_BO}
\end{equation}
and the extent of the Bloch oscillations
\begin{equation}
L_{\mathrm{BO}}=\frac{4J}{F_{0}}.
\end{equation}

\subsection{Bloch translation}
\label{sec.dyn.BT}

In the following, we will revisit the dynamics of a wave packet
subjected to a periodic time-modulated field $F(t)$ with the aid of
the propagator, which offers advantages over the traditional
approach. First, let us consider the external field in the form
\begin{equation}
F\left(t\right) =n\omega +F_{\mathrm{A}}\cos \left( \omega t\right),
\end{equation}
where $n$ is an integer representing the DC field, and
$F_{\mathrm{A}}$ is a constant representing the AC field. According
to our former analysis, we have
\begin{equation}
I \left( t \right) = n \omega t + \frac{F_{\mathrm{A}}}{\omega} \sin
\left( \omega t \right) ,
\end{equation}
and
\begin{equation}
v_{\mathrm{g}}\left( t\right) =2J\sin \left[ k_{0}-n\omega t-\frac{F_{%
\mathrm{A}}}{\omega }\sin \left( \omega t\right) \right] .
\label{eq.BT.vg}
\end{equation}

\begin{figure}[tbp]
\begin{center}
\includegraphics[bb=0 0 586 469,width=0.45\textwidth,clip]%
{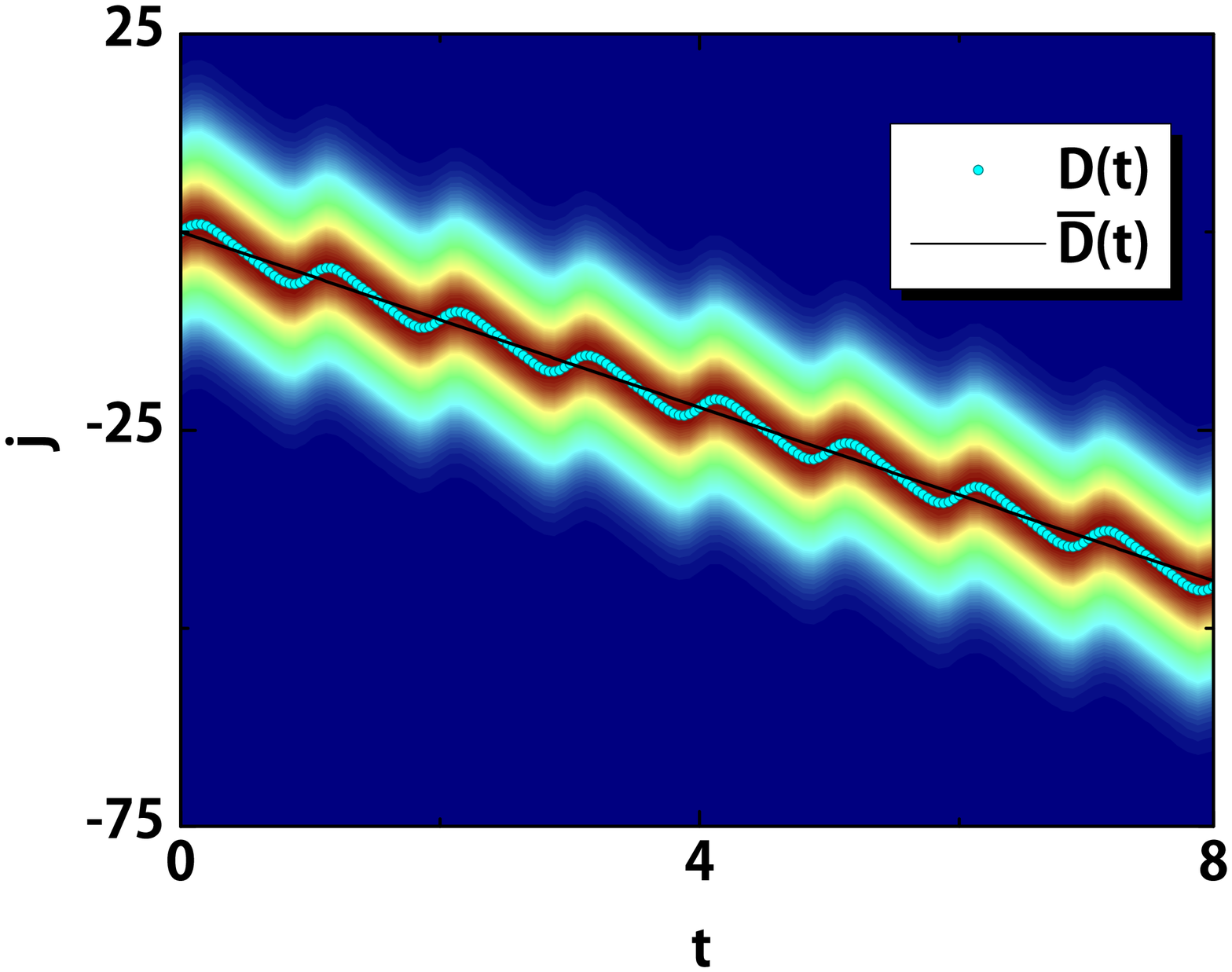}
\includegraphics[bb=0 0 586 470,width=0.44\textwidth,clip]%
{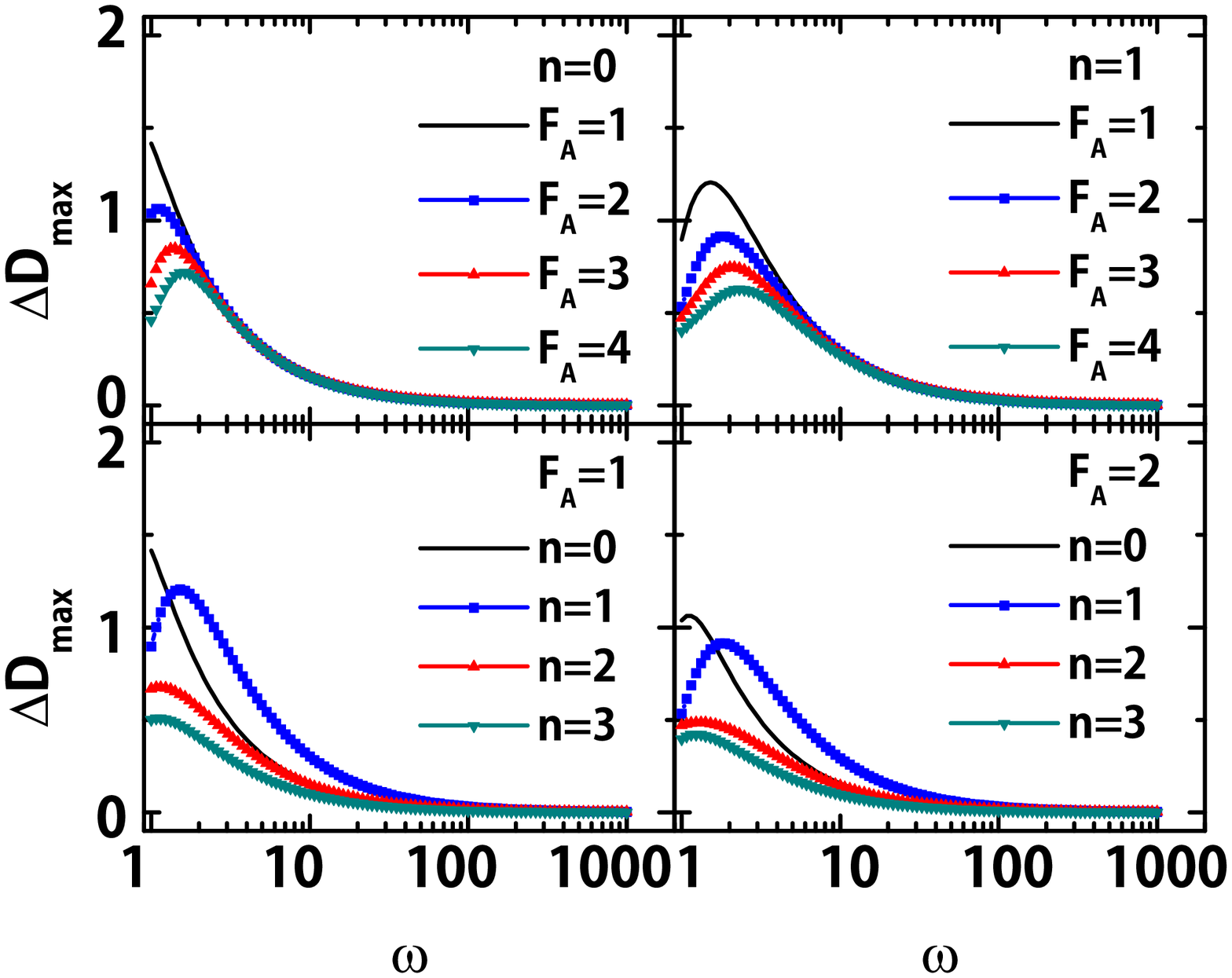}
\caption{(Color online) The time evolutions of a Gaussian wave packet
(GWP) are plotted of the envelopes with time $t$ (left) and the
shaking deviations with the field frequency $\omega$ (right). The
envelopes are calculated by exact diagonalization from
(\ref{eq.U_div}), where the external field is $n = 1$,
$F_{\mathrm{A}} = 1$ and $\omega = 1$. The displacement (\ref{Dt_C})
and the average displacement (\ref{eq.BT.DtBar}) are marked by the
solid circle and line in units of the lattice site, where $t$ is in
units of $\tau = 2 \pi / \omega$. The maximum shaking amplitudes
$\Delta D _{\mathrm{max}}$ from (\ref{eq.BT.DDt}) of the same wave
packet are plotted under the fields with different $n$,
$F _{\mathrm{A}}$ (in units of $J$) and $\omega$ (in units of $J$).
The wave packet from (\ref{eq.GWP}) has $\alpha=0.1$ and central
momentum $k_0 = \pi /2$ on an $N = 200$ chain with hopping amplitude
$J=1$. It shows that the GWP moves straightly with small shaking,
which is depressed under high frequency external field.}
\label{fig.BT}
\end{center}
\end{figure}

We notice that $v_{\mathrm{g}}\left( t\right) $ is a periodic
function with the period $\tau =2\pi /\omega $, but the displacement
in each period
\begin{eqnarray}
D\left( \tau \right)
&=& \int_{0}^{\tau } v_{\mathrm{g}} \left( t \right) \rmd t
\nonumber \\
&=& 2J \left( -1 \right)^{n} \mathcal{J}_{n} \left(
\frac{F_{\mathrm{A}}}{\omega}\right) \sin \left( k_{0} \right) \tau
\label{D_tau_BT}
\end{eqnarray}
is non-zero in general, since $v_{\mathrm{g}}\left( t\right) $ of
(\ref{eq.BT.vg}) is not a monochromatic function of time $t$. Here
$\mathcal{J}_{n} \left( z \right)$ denotes the Bessel function of the
first kind. Then, the wave packet exhibits unidirectional motion with
a periodic shaking, which can be referred to as the Bloch
translation.

Considering a long time scale of $t \gg \tau $, we have
\begin{equation}
\overline{D}\left( t\right) \approx 2J \left( -1 \right)^{n}
\mathcal{J}_{n} \left( \frac{F_{\mathrm{A}}}{\omega } \right) \sin
\left( k_{0} \right) t, \label{eq.BT.DtBar}
\end{equation}
which indicates the aperiodicity of the displacement. The wave packet
evolves as if in a uniform tight-binding chain, with the effective
hopping amplitude $J_{\mathrm{eff}} = J \left\vert \mathcal{J}_{n}
\left( F_{\mathrm{A}} / \omega \right) \right\vert$. Meanwhile, the
deviation of the central position from its average
\begin{equation}
\Delta D=D\left( t\right) -\overline{D}\left( t\right)
\label{eq.BT.DDt}
\end{equation}
is also a periodic function
\begin{equation}
\Delta D\left( t + \tau \right) =\Delta D\left( t\right) .
\end{equation}
Then, the wave packet moves in one direction accompanied by small
shaking. From (\ref{eq.BT.vg}), the instantaneous velocity
$v_{\mathrm{g}} \left( t \right)$ is a bounded function with
$\left\vert v_{\mathrm{g}} \left( t \right) \right\vert \leq 2J$. And
we have
\begin{equation}
\lim_{\omega \rightarrow \infty }\Delta D\left( t\right) =0,
\end{equation}
because the period $\tau$ goes to infinitesimal. This indicates that
the shaking disappears as the frequency increases, which is shown in
Figure~\ref{fig.BT}. Then, the wave packet moves uniformly with a
constant velocity $2J_{\mathrm{eff}} \sin \left( k_0 \right)$.

In principle, the wave packet can be coherently stopped when an
appropriate $F_{\mathrm{A}}$ is taken to make $\mathcal{J}_{n}\left(
F_{\mathrm{A}} / \omega \right) =0$. We will consider this problem
for the two cases $n=0$ and $n\geq 1$. For the zeroth-order Bessel
function of the first kind, we have $\mathcal{J}_{0} \left(
F_{\mathrm{A}} /\omega \right) =0$ when $F_{ \mathrm{A}} / \omega =
0.768 \pi$, etc., which are the roots of the Bessel function. It
indicates that one should take frequency-dependent $F_{\mathrm{A}}$
to obtain DD. As for the ring system with the magnetic flux as a
monochromatic wave of
\begin{equation}
\phi \left( t\right) =\phi _{\mathrm{A}}\sin \left( \omega t\right) ,
\end{equation}
it corresponds to $\phi _{\mathrm{A}} = - F_{\mathrm{A}} / \omega =
-0.768 \pi$, which accords with the results in our previous study
\cite{Hu_AC}. Nevertheless, in the case of $n \geq 1$, it becomes
simple since the property of the Bessel function makes
$\mathcal{J}_{n} \left( 0 \right) =0$ for $ n \geq 1$. Thus, in such
a case, DD can be realized for any finite value of $F_{\mathrm{A}}$
in the limit of $\omega \rightarrow \infty $.

\subsection{Super Bloch oscillations}
\label{sec.dyn.SBO}

Now, we consider the external field in the form of
\begin{equation}
F \left( t \right) = \left( n + \delta \right) \omega +
F_{\mathrm{A}} \cos \left( \omega t\right) ,
\end{equation}
where $n$ is an integer, and $\delta$ is the detuning factor. Such a
model has been studied extensively by the equation of motion for the
average of the observables. In this study, we revisit it by
investigating the time evolution of the wave function. Our former
analysis gives
\begin{equation}
I \left( t \right) = n \omega t + \delta \omega t +
\frac{F_{\mathrm{A}}}{\omega} \cos \left( \omega t \right),
\end{equation}
and
\begin{equation}
v_{\mathrm{g}} \left( t \right) = 2J\sin \left[ k_{0} - n \omega t -
\delta \omega t - \frac{F_{\mathrm{A}}}{\omega } \sin \left( \omega
t \right) \right] . \label{eq.SBO.vg}
\end{equation}
In this case, $v_{\mathrm{g}} \left( t \right) $ is no longer a
periodic function of time $t$ with the period $\tau$. Because of the
slight detuning of $\delta \ll 1$, it can be presumed that the group
velocity $v_{\mathrm{g}} \left( t \right) $ still oscillates with a
quasi-period $\tau$. However, the displacement during each
quasi-period is not invariable and can change its sign at a certain
instant. More precisely, the one-period displacement is given by
(\ref{eq.BT.DtBar}) as
\begin{equation}
D\left( \tau \right) \approx 2J\left( -1\right) ^{n} \mathcal{J}_{n}
\left( \frac{F_{\mathrm{A}}}{\omega } \right) \sin \left( k_{0} -
\delta \omega t\right) \tau ,
\end{equation}
which is a periodic function with a long period
\begin{equation}
\tau _{\mathrm{SBO}}=\frac{2 \pi}{\delta \omega} \label{eq.tau_SBO}
\end{equation}
approximately. Accordingly, the one-period average displacement has
the form
\begin{eqnarray}
\overline{D}\left( t\right)
&=& \sum_{l=1}^{\mathrm{Int} \left( t/ \tau \right)} \int_{\left( l
- 1 \right) \tau }^{ l \tau } v_{\mathrm{g}} \left( t^{\prime }
\right) \rmd t^{\prime }  \nonumber \\
&\approx& 2J \frac{\left( -1 \right)^{n}}{\delta \omega }
\mathcal{J}_{n} \left( \frac{F_{\mathrm{A}}}{\omega} \right) \left[
\cos \left( k_{0} - \delta \omega t \right) - \cos \left( k_{0}
\right) \right] ,  \label{D_ave}
\end{eqnarray}
where $\mathrm{Int}$ denotes the greatest integer function.

\begin{figure}[tbp]
\begin{center}
\includegraphics[bb=0 0 586 469,width=0.45\textwidth,clip]%
{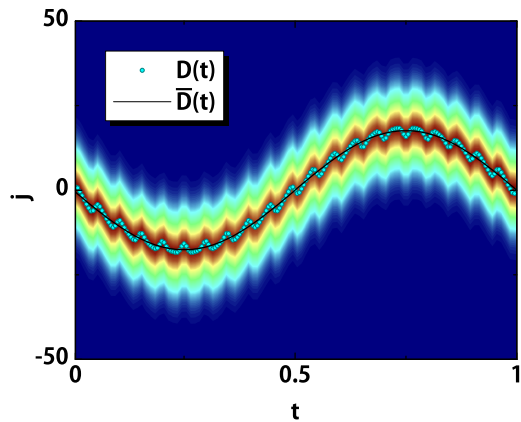}
\caption{(Color online) The envelope of the same GWP on the same
chain as shown in the left of Figure~\ref{fig.BT}, while with a
detuning $\delta = 0.02$ of the external field and $t$ in units of
$\tau_{\mathrm{SBO}}$ from (\ref{eq.tau_SBO}). It shows that the GWP
behaves super Bloch oscillations with large extent and small
shaking.}
\label{fig.SBO}
\end{center}
\end{figure}

We can see that $\overline{D}\left( t\right) $ has the same form of
the customary Bloch oscillations as (\ref{D_BO}) in an infinite chain
system with the following effective hopping amplitude and the static
force of
\begin{eqnarray}
J_{\mathrm{eff}} &=& J \left\vert \mathcal{J}_{n} \left(
\frac{F_{\mathrm{A}}}{\omega} \right) \right\vert , \\
F_{\mathrm{eff}} &=& \delta \omega .
\end{eqnarray}
This has been termed super Bloch oscillations with the extent
\begin{equation}
L_{\mathrm{SBO}} = \frac{4J_{\mathrm{eff}}}{F_{\mathrm{eff}}}
= 4J \left\vert \frac{\mathcal{J}_{n} \left( \frac{F_{\mathrm{A}}}{
\omega} \right) }{\delta \omega} \right\vert.
\end{equation}
We can get the average group velocity as
\begin{eqnarray}
\overline{v}_{\mathrm{g}}\left( t\right)
&=& \frac{\partial \overline{D}\left( t\right) }{\partial t}
\nonumber \\
&=& -2J \left( -1 \right)^{n} \mathcal{J}_{n} \left(
\frac{F_{\mathrm{A}}}{\omega} \right) \sin \left( k_{0} - \delta
\omega t\right) .
\end{eqnarray}
Obviously, $\overline{v}_{\mathrm{g}} \left( t\right) $ is periodic
with the period $2 \pi / \delta \omega$, which is the period of the
super Bloch oscillations. To illustrate our analysis,
Figure~\ref{fig.SBO} shows plots of $D\left( t \right) $ in
(\ref{Dt_C}) and $\overline{D}\left( t\right)$ in (\ref{D_ave}) as
the functions of time $t$ in units of $\tau_{\mathrm{SBO}}$. For
comparison, we also plot the profile of the time evolution of a wave
packet, which is directly computed from (\ref{eq.U_div}) by the exact
diagonalization of the finite-size Hamiltonians $H\left( n \epsilon
\right) $ with $\epsilon =0.01/J$. This shows that our analysis is in
agreement with the numerical simulation and provides a more exact
picture of super Bloch oscillations.

\section{Quantum state manipulation}
\label{sec.QST}

Coherent quantum state storage and transfer via a coupled qubit
system are crucial in the emerging area of QIP. The essence of such a
process is the time evolution of a local wave packet in a discrete
system. In the design of the scheme for quantum state transfer, two
types of external control are usually employed: one adiabatic and the
other requiring sudden changes of parameters. In the first scheme,
the external field is turned on and off in an adiabatic manner. Then,
the quantum state is fixed on the superposition of the instantaneous
eigenstates. This requires a sufficiently long relaxation time to
satisfy the adiabatic approximation condition. Therefore, decoherence
becomes incredibly challenging for this scheme to be realized. The
second scheme requires a stepwise change of the external field, which
is difficult to accomplish practically. The scheme that employs the
diabatic process shows more promise. The main difficulty with
accomplishing this task in a quantum network is the tricky issue in
quantum mechanics, which is the time evolution of a time-dependent
Hamiltonian.

\begin{figure}[tbp]
\begin{center}
\includegraphics[bb=0 0 586 532,width=0.45\textwidth,clip]%
{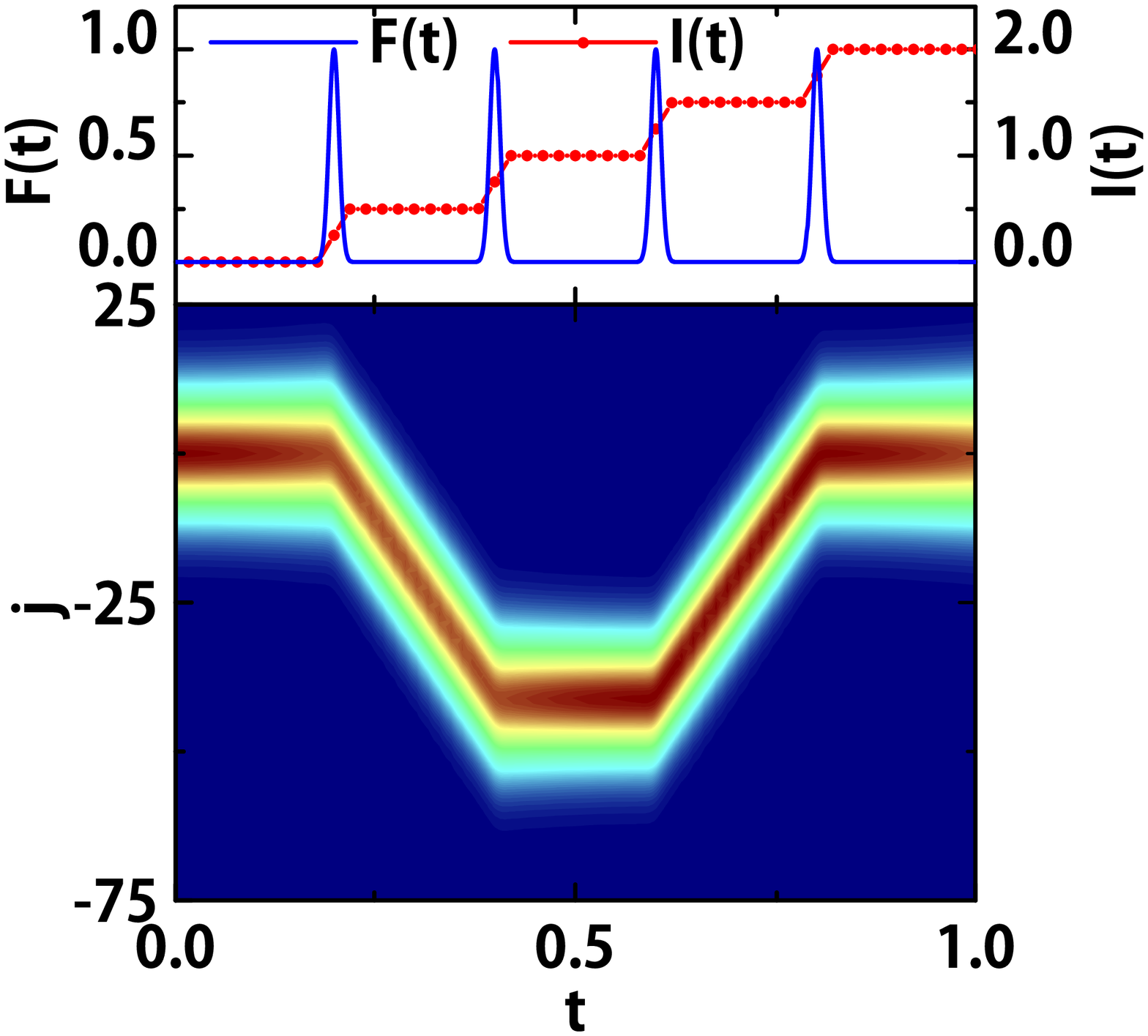}
\includegraphics[bb=0 0 586 536,width=0.45\textwidth,clip]%
{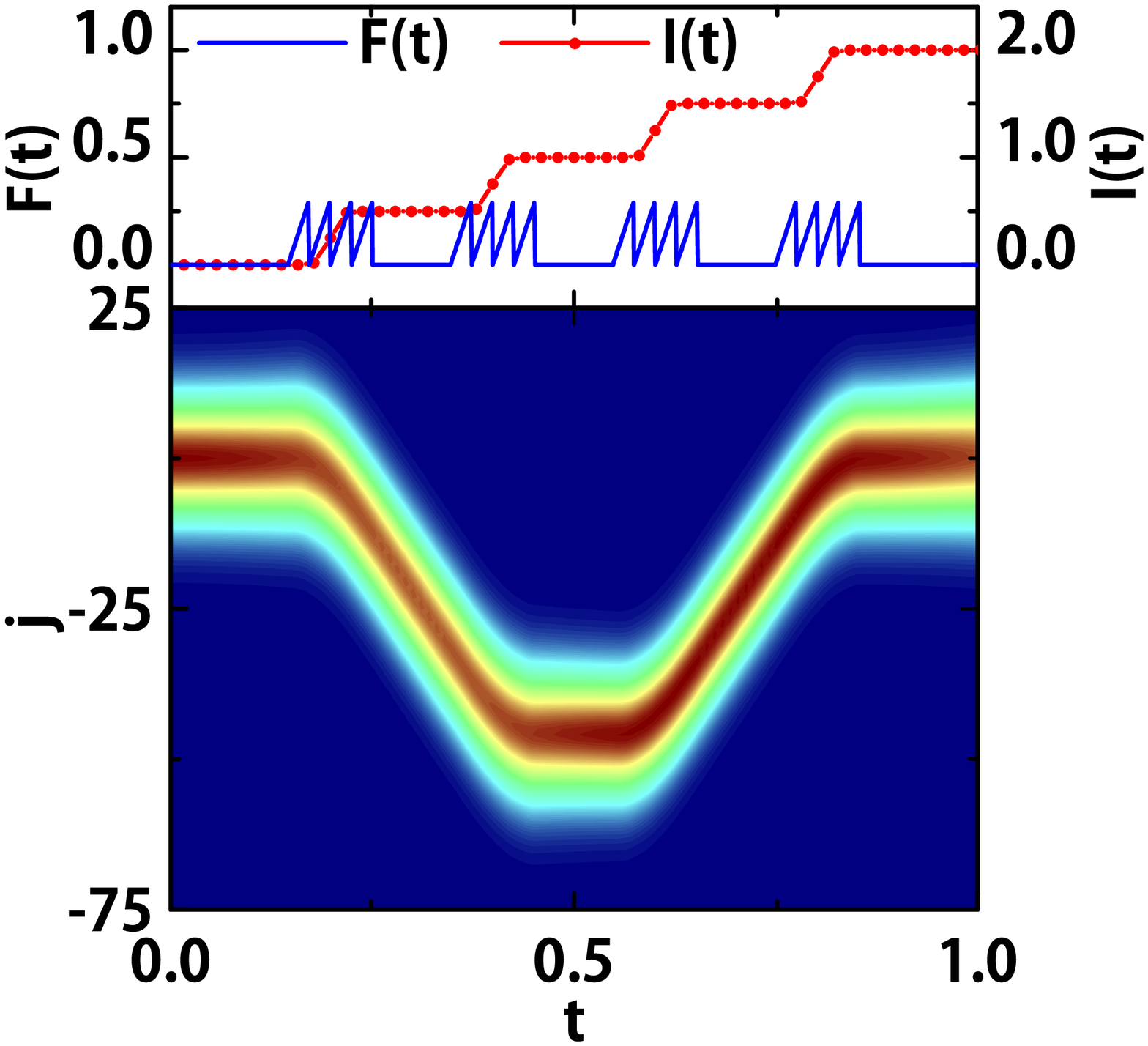}
\caption{(Color online) The envelopes of the time evolution of
same Gaussian wave packet (GWP) on the same chain as that in
Figure~\ref{fig.BT}, with different trains of pulsed field $F\left( t
\right)$ which are Gaussian (left) and sawtooth-like (right), where
$j$ is in units of the lattice site, $F \left( t \right)$ is in units
of $J$, the impulse of the train accumulation $I \left( t \right)$ is
in units of $\pi$, and $t$ is in units of $N / 2J$. We can see that
the group velocity is changed at the end of each pulse. The dynamics
under the two trains has the same behaviour despite the details
during the changes.}
\label{fig.QST}
\end{center}
\end{figure}

According to our former analysis, the analytical evolved wave
function in the present two systems can be obtained for an arbitrary
time-dependent external field. Furthermore, it reveals an interesting
feature of the wave packet dynamics: the wave packet preserves its
shape for an arbitrary profile of the time-modulated external field
$F(t)$ owing to the homogeneity of force in space, while the momentum
of the wave packet changes according to the impulse-momentum theorem.
This enables us to propose such a scheme, in which the adiabaticity
of the tuning process is not required. We illustrate this point as
follows.

We consider the linear potential as being a Gaussian-pulse train,
which can be expressed as
\begin{eqnarray}
F \left( t \right) &=& \sum_{n} F_{n} \left( t\right), \\
F_{n} \left( t\right) &=& \frac{\sqrt{\pi}}{2\sigma} \exp \left[ -
\frac{ \left( t-T_{n} \right) ^{2} }{\sigma ^{2}} \right],
\end{eqnarray}
where $\sigma$ determines the duration of each pulse. In the case
$T_{n+1}-T_{n} \gg \sigma $, we have the impulse as
\begin{equation}
I_{n}=\frac{\sqrt{\pi }}{2\sigma }\int_{-\infty }^{\infty } F_{n}
\left( t \right) \rmd t=\frac{\pi }{2}.
\end{equation}
Here we choose the impulse to be $\pi /2$ since it is the minimal
momentum difference between the static and the fastest-moving wave
packets. According to the above results, a GWP acquires an extra
momentum $\pi /2$ after each pulsed field. For an initial GWP with
$k_{0}=0$, $I_{1}$ should accelerate it to the momentum of $\pi/2$
with the speed of $2J$, and $I_{2}$ should shift its momentum to
$\pi$, i.e. stops it. The next field pulse $I_{3}$ turns it back and
the final pulse $I_{4}$ stops it. This behaviour is illustrated in
Figure~\ref{fig.QST}, which shows the time evolution for an initially
motionless GWP in a field composed by a sequence of four identical
pulses with $\sigma =0.886$ and $t_{n}=nN/10J$, $n=1,2,3$ and $4$.
The time evolution is directly computed from (\ref{eq.U_div}) by
exact diagonalization of the finite-sized Hamiltonians
$H\left( n\epsilon \right) $ with $n=\left[ 0,1.7\times 10^{4}
\right]$ and $\epsilon =5.9 \times 10^{-3}/J$. The speed of the wave
packet is changed at the end of each pulse. This behaviour
illustrates the expected phenomena of accelerating, stopping and
turning of the wave packet.

Furthermore, if the external pulsed field is not Gaussian-like, but
arbitrarily shaped with the same impulse (the area of the pulse), the
result should be the same. To demonstrate this, we compute the time
evolution for a sawtooth-pulsed train in Figure~\ref{fig.QST}. We can
see that despite the turning points, the different pulse train gives
the same result in controlling the wave packet, and because of the
accumulation of impulse, the amplitude of the external field can be
lowered by extending the width of the pulses in the train. When being
applied along with the control scheme, this coherent accelerating
stopping turning scheme provides a useful strategy for the efficient
coherent transfer of a quantum state.

\section{Summary}
\label{sec.Sum}

In conclusion, the exact propagators of these two one-dimensional
systems have shown that the Bloch acceleration theorem can be
generalized to the impulse-momentum theorem in a quantum version,
which connects the impulse not only to the shift of the momentum but
also to the phase change of the matter waves. It has been
theoretically shown that accelerating, stopping coherently and
transferring a wave packet to a location on demand can be realized by
exploiting the time-dependent external field. The proposed scheme is
less distinct than the adiabatic process and provides a noteworthy
example of coherent quantum control with high fault tolerance. It
offers an advantage in that neither the precise modulation of field
pulses nor adiabatic conditions for the dynamical field changes are
necessarily required.

\ack We acknowledge the support of National Basic
Research Program (973 Program) of China under Grant No.~2012CB921900.

\end{document}